\documentclass[final,5p,authoryear,twocolumn]{elsarticle}

\usepackage{amssymb,amsmath}
\usepackage{braket}
\usepackage{pdflscape}

\usepackage{lineno}

\journal{arXiv}
\newcommand{\ul}{\textbf}

\begin{document}

\begin{frontmatter}



\title{Cooperative ``folding transition'' in the sequence space facilitates function-driven evolution of protein families}


\author{Akira R. Kinjo}
\ead{akinjo@protein.osaka-u.ac.jp}
\address{Institute for Protein Research, Osaka University, 3-2 Yamadaoka, Suita, Osaka, 565-0871, Japan}

\begin{abstract}
In the protein sequence space, natural proteins form clusters of families which are characterized by their unique native folds whereas the great majority of random polypeptides are neither clustered nor foldable to unique structures.
Since a given polypeptide can be either foldable or unfoldable, a kind of ``folding transition'' is expected at the boundary of a protein family in the sequence space. By Monte Carlo simulations of a statistical mechanical model of protein sequence alignment that coherently incorporates both short-range and long-range interactions as well as variable-length insertions to reproduce the statistics of the multiple sequence alignment of a given protein family, we demonstrate the existence of such transition between natural-like sequences and random sequences in the sequence subspaces for 15 domain families of various folds.
The transition was found to be highly cooperative and two-state-like.
Furthermore, enforcing or suppressing consensus residues on a few of the well-conserved sites enhanced or diminished, respectively, the natural-like pattern formation over the entire sequence. In most families, the key sites included ligand binding sites. These results suggest some selective pressure on the key residues, such as ligand binding activity, may cooperatively facilitate the emergence of a protein family during evolution. From a more practical aspect, the present results highlight an essential role of long-range effects in precisely defining protein families, which are absent in conventional sequence models.
\end{abstract}


\end{frontmatter}


\section{Introduction}
Natural proteins can be classified into families based on their sequence similarity\citep{Pfam}. This is considered to be primarily a consequence of molecular evolution: proteins evolved from a common ancestral protein share similar sequences. However, evolution alone does not account for the existence of relatively well-defined (domain) families that are distributed rather discretely than continuously in the sequence space \citep{Maynard-Smith1970,Nishikawa1993,Nishikawa2002,Goldstein2008}. A key to understanding the family distribution is protein folding. As observed in protein structure classification databases \citep{SCOP,CATH,ECOD}, each protein family corresponds to a unique three-dimensional fold, suggesting the existence of physical constraints imposed on protein sequences during the evolutionary process to maintain the fold \citep{MorcosETAL2014}. While protein structures can tolerate great many mutations to the extent that proteins with little sequence similarity can share the same fold, residue conservation patterns reflect the structural context of protein sequences. This fact has long been exploited in protein structure prediction in the form of position-specific scoring matrices \citep{Taylor1986,GribskovETAL1987,AltschulETAL1997,KinjoANDNakamura2008} and, more recently, direct-coupling analysis and related methods \citep{BalakrishnanETAL2011,MorcosETAL2011,JonesETAL2012,TaylorETAL2012,Miyazawa2013,EkebergETAL2013,Kinjo2015,Levy2017}.

Under a given physiological condition, a polypeptide is either able or unable to fold into some unique structure. This suggests the existence of a ``folding transition'' at the border between an ``island'' of a protein family and the ``sea'' of random polypeptide, that is analogous to the folding transition of a protein molecule in the conformational space \citep{Nishikawa1993,Nishikawa2002,ShakhnovichANDGutin1993b}. It should be noted, however, that there are many families in the sequence space so that a sequence moving in the sequence space may fall into any one of these families. This is in contrast to protein folding in the conformational space where there is usually only one unique native structure for a given protein sequence. Furthermore, a (structural) domain, rather than a whole protein sequence, should be considered as a unit of folding as a particular domain may be found in different proteins in combination with other, different, domains.
Therefore, the system in which the analogy of protein folding holds should be limited to the vicinity of each protein domain family rather than the entire sequence space. In the following, we focus on the folding transition in a sequence subspace around a given protein domain family. Although biologically important, intrinsically disordered proteins \citep{DunkerETAL2001,MinezakiETAL2006,Tompa2012} are excluded from the present study for the following two reasons. First, the analogy of the folding transition may not apply to those proteins. Second, it is difficult to obtain reliable and comprehensive multiple sequence alignments for this class of proteins \citep{LangeETAL2015}, which are required for parameter estimation of the statistical model employed in the present study. 

There have been a number of theoretical and computational studies on subjects related to the sequence space such as foldability and design \citep{ShakhnovichANDGutin1993,ShakhnovichANDGutin1993b,GovindarajanANDGoldstein1995,MorcosETAL2014}, molecular evolution within an island \citep{Bornberg-BauerANDChan1999,BastollaETAL1999,WroeETAL2005} and between islands \citep{WroeETAL2007,HolzgrafeANDWallin2014,SikosekETAL2016}, or the size and/or distribution of islands in the sequence space \citep{GovindarajanANDGoldstein1996,LiETAL1996,Bornberg-Bauer1997,KuhlmanANDBaker2000,KoehlANDLevitt2002}.
On the contrary, relatively little attention has been paid to the transition between an island (a set of sequences belonging to the same family) and the sea (the set of sequences that do not belong to the family) apart from a few exceptions. In the context of protein design, \citet{ShakhnovichANDGutin1993b} theoretically predicted the existence of a ``folding transition''.  In a study of hierarchical evolution of protein fold families based on a simple model of the evolutionary selection by native stability using a generic contact potential \citep{MiyazawaANDJernigan1985}, \citet{DokholyanANDShakhnovich2001} observed a sharp transition at a certain design temperature. In neither of these studies, however, the nature of the transition was investigated further.
Characterizing the folding transition in the sequence space may help understand essential features that constitute a protein family and possible evolutionary trajectories that may have led to the emergence of a protein family. It also has practical importance in identifying new family members and designing new proteins. 

In the following, we investigate the folding transition in the sequence subspaces for 15 protein domain families including all-$\alpha$, all-$\beta$, $\alpha/\beta$ and other folds by performing extensive Monte Carlo (MC) simulations of the modified lattice gas model (LGM) of protein sequence alignment \citep{Kinjo2016,Kinjo2017} that coherently integrates long-range interactions and variable-length insertions. Using the LGM, the existence of a sharp two-state transition between natural-like sequences and random sequences in the sequence subspace is demonstrated. Furthermore, the nature of the transition is examined in detail by analyzing residue distribution of each site along the transition as well as by performing virtual ``mutation'' experiments.

\section{Theory}
\begin{figure}
\includegraphics[width=0.45\textwidth]{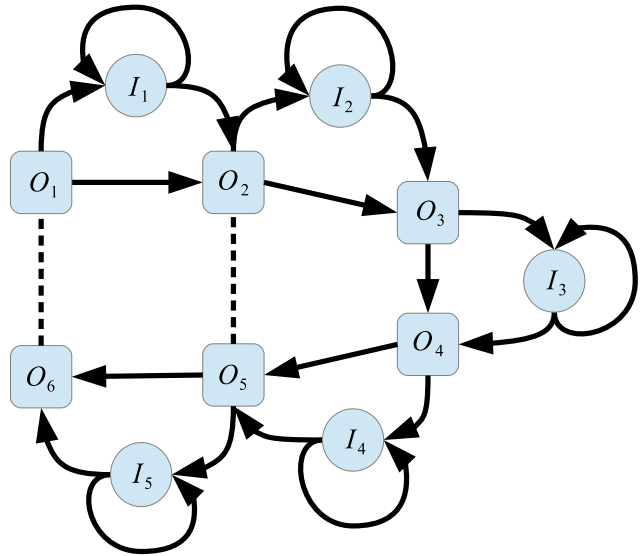}
\caption{\label{fig:model} An example model structure with model length $N=6$. There are $N$ core sites $O_1,\cdots, O_N$ and $N-1$ insert sites $I_1,\cdots, I_{N-1}$.
These model sites are bonded via ``bonds'' (solid arrows). Some pairs of non-bonded core sites may be ``interacting'' (dashed lines). }
\end{figure}

We briefly summarize the theory to the extent that is necessary for understanding the present study. For more details about the formulation of the LGM as well as the algorithms for MC simulations and parameter optimization, refer to the previous papers \citep{Kinjo2016,Kinjo2017}.
The LGM for a given Pfam \citep{Pfam} family consists of $N$ ``core'' sites and $N-1$ ``insert'' sites which respectively correspond to the ``match'' states and ``insert'' states of the Pfam profile hidden Markov model (HMM) of length $N$,
excluding the N- and C-terminal insert states. Exactly one of the 21 residue types (including the ``delete'' symbol) can exist on each core site whereas arbitrarily many (including zero) residues out of the standard 20 residue types can reside at each insert site.
The array of core and insert sites are connected via ``bonds'' (solid arrows in Fig. \ref{fig:model}) that reflect the linear polypeptide structure. A pair of model sites connected via a bond are called a ``bonded pair'' in the following. Between core sites more than 2 residues apart along the sequence, there may be interactions (dashed lines in Fig. \ref{fig:model}) based on a representative native structure of the family. Two sites are defined to be interacting if the residues aligned to those sites are in contact in the corresponding representative native structure. Two residues are defined to be in contact if any non-hydrogen atoms in those residues are within 5\AA. Interactions are defined only between core sites for simplicity. Interacting core sites are referred to as ``non-bonded'' pairs in the following.

Let $\mathbf{a} = a_1\cdots a_L$ be an amino acid sequence of $L$ residues and an LGM $\mathcal{M}$ of length $N$ consist of core sites $O_1,\cdots, O_N$ and insert sites $I_1 ,\cdots, I_{N-1}$. An alignment between the sequence $\mathbf{a}$ and the model $\mathcal{M}$ is represented as a sequence of pairs of a model site (core or insert) and a residue: $\mathbf{X} = X_1\cdots X_{L_{\mathbf{X}}}$ where
$L_{\mathbf{X}}$ is the length of the alignment and each $X_i$ is a pair such as $(S,a)$ with $S\in \{O_i\}_{i=1,\cdots,N} \cup \{I_i\}_{i=1,\cdots,N-1}$ and $a \in \{a_1,\cdots, a_L\}$.
For example, given an amino acid sequence, say \texttt{KCFPDGVW}, and a model of length $N=6$ (Fig. \ref{fig:model}), one of many possible alignments is represented as
$\mathbf{X} = X_{1}\cdots X_{9} =$
$(O_1,\mathtt{K})$
$(O_2,\mathtt{C})$
$(O_3,\mathtt{F})$
$(I_3,\mathtt{P})$
$(I_3,\mathtt{D})$
$(I_3,\mathtt{G})$
$(O_4,\mathtt{-})$
$(O_5,\mathtt{V})$
$(O_6,\mathtt{W})$. Note there are multiple occurrence of the insert site $I_3$ whereas other insert sites are completely absent in this particular alignment.
Since there may be any number of residues at each insert site, the alignment length is variable.

Based on this representation of sequence alignment, the energy function of alignment $\mathbf{X}$ is defined as
\begin{eqnarray}
  E(\mathbf{X}) &=&
  -{\sum_{k=1}^{L_{\mathbf{X}}-1}}J(X_k,X_{k+1})
  -{\sum_{(k,l) \in \mathcal{T}}}K(X_k,X_l) \nonumber\\
&&  -{\sum_{k=1}^{L_{\mathbf{X}}}}\mu(X_k)
    \label{eq:energy0}
\end{eqnarray}
where $J$ and $K$ are short-range and long-range interaction parameters, respectively, $\mu$'s are chemical potentials, and $\mathcal{T}$ indicates the set of all the interacting non-bonded pairs. The short-range interactions act only between bonded pairs of residues that are consecutive in the alignment (i.e., between $X_k$ and $X_{k+1}$ in Eq. \ref{eq:energy0}). The long-range interactions act between residues that are
aligned to interacting non-bonded pairs. The chemical potentials are so called because they are used to control the residue densities of each site. Only $J$ and $K$ parameters constitute intrinsic energy, and they are to be determined from a given (observed) multiple sequence alignment (MSA) of the family sequences (see below).

We assume that the probability $P(\mathbf{X})$ of obtaining an alignment $\mathbf{X}$ is given by the Boltzmann distribution:
\begin{equation}
  \label{eq:boltzmann}
  P(\mathbf{X}) = \frac{\exp[-E(\mathbf{X})/T]}{\Xi[T]}
\end{equation}
where $T$ is the ``design (selection) temperature''\citep{ShakhnovichANDGutin1993,ShakhnovichANDGutin1993b} in energy unit and $\Xi[T]$ is the partition function 
\begin{equation}
\Xi[T] = \sum_{\mathbf{X}}\exp[-E(\mathbf{X})/T]\label{eq:partition-function}.
\end{equation}
Here, the summation is over all possible alignments ($\mathbf{X}$) of all possible amino acid sequences with the model. Since the alignment length can vary, this ensemble is considered to be a grand canonical ensemble.

Given the parameters $J$, $K$ and $\mu$, we can sample (alignments of) sequences according to the probability distribution Eq. (\ref{eq:boltzmann}) by running grand canonical Monte Carlo (MC) simulations \citep{Kinjo2017}.

We define the \emph{standard condition} as the system with $\mu(X_k) = 0$ for all $X_k$ (i.e., no perturbation), and the \emph{natural condition} as the standard condition with $T = 1$. 
The parameters $J$ and $K$ are determined iteratively so that the pair number densities of residues for bonded and non-bonded pairs over the samples produced by MC simulations under the natural condition match those observed in the given MSA \citep{LapedesETAL1999,SuttoETAL2015}. This optimization corresponds to minimizing the following free energy function under the natural condition ($T=1$ in particular):
\begin{equation}
  F[T] = -\max_{J,K}(\braket{E}_{\mathrm{obs}} - \Omega[T])\label{eq:fe}
\end{equation}
where $\braket{E}_{\mathrm{obs}}$ is the average energy of the observed MSA and $\Omega[T] = -T\ln\Xi[T]$ is the grand potential. It can be easily shown that this is equivalent to the principle of maximum entropy \citep{LapedesETAL1999,MorcosETAL2011}.

The design temperature in the LGM controls the mutation rate uniformly over the entire sites. The higher the temperature, the higher the mutation rate. The temperature of the natural condition $T=1$ \emph{defines} the unit of energy and temperature. In other words, the temperature for each domain family is in a relative unit defined by the natural condition (unlike the absolute scale as in \citet{MorcosETAL2014}). The mutation rate here means the number of accepted mutations per attempted mutation, and mutations are attempted at a constant time interval. The relationship between the present ``energy'' and physical folding free energy has been recently established by \citet{Miyazawa2017}.

\section{Results}
\begin{table*}
  \caption{\label{tab:families}
    \small
    Pfam domain families used in the study.}
    \begin{tabular}{lp{3cm}rlrlrr}
      \hline
      Pfam ID & description & $N^a$  & MSA$^b$ & $M^c$ & PDB$^d$ & $T_D$$^e$ & $T_m$$^f$\\\hline
      zf-C2H2 &  Zinc finger, C2H2 type & 23 & uniprot & 503700 & 4m9vC & 1.085 & 1.033 \\ 
SH3\_1 &  SH3 domain & 48 & uniprot & 30001 & 4hvuA & 1.050 & 1.019 \\ 
HTH\_3 &  Helix-turn-helix & 55 & uniprot & 97452 & 1r69A & 1.045 & 1.019 \\ 
Homeobox &  Homeobox domain & 57 & uniprot & 47752 & 4rduD & 1.040 & 1.015 \\ 
HisKA &  His Kinase A (phospho-acceptor) domain & 64 & uniprot & 224625 & 2c2aA & 1.045 & 1.018 \\ 
RRM\_1 &  RNA recognition motif. (a.k.a. RRM, RBD, or RNP domain) & 70 & uniprot & 124098 & 1l3kA & 1.045 & 1.018 \\ 
PDZ &  PDZ domain (Also known as DHR or GLGF) & 82 & uniprot & 34441 & 1g9oA & 1.035 & 1.013 \\ 
fn3 &  Fibronectin type III domain & 85 & uniprot & 104374 & 2ic2A & 1.050 & 1.019 \\ 
I-set &  Immunoglobulin I-set domain & 90 & uniprot & 93895 & 1u2hA& 1.035 & 1.014 \\ 
OmpA &  OmpA family & 95 & uniprot & 44148 & 4zhwA & 1.025 & 1.010 \\ 
DnaB &  DnaB-like helicase N terminal domain & 103 & uniprot & 14624 & 2r5uA & 1.030 & 1.009 \\ 
V-set &  Immunoglobulin V-set domain & 109 & uniprot & 10392 & 4unuA & 1.035 & 1.012 \\ 
Globin &  Globin & 110 & uniprot & 6055 & 2nrlA & 1.030 & 1.011 \\ 
HATPase\_c &  Histidine kinase-, DNA gyrase B-, and HSP90-like ATPase & 111 & rp55 & 106253 & 4xe2A  & 1.030 & 1.011 \\ 
Response\_reg &  Response regulator receiver domain & 112 & full & 98600 & 3chyA & 1.025 & 1.010 \\ 
\hline
    \end{tabular}\\
    {\small
    $^a$Length of Pfam profile HMM.
    $^b$Dataset for Pfam multiple sequence alignment.
    $^c$Number of sequences in the MSA with less than 10\% deletes.
    $^d$PDB \citep{wwPDB} chains used for computing three-dimensional contacts. References:
    4m9v \citep{PDB4M9V};  4hvu \citep{PDB4HVU}; 1r69 \citep{PDB1R69}; 4rdu \citep{PDB4RDU};
    2c2a \citep{PDB2C2A}; 1l3k \citep{PDB1L3K}; 1g9o \citep{PDB1G9O};  2ic2 \citep{PDB2IC2};
    1u2h \citep{PDB1U2H}; 4zhw \citep{PDB4ZHW}; 2r5u \citep{PDB2R5U}; 4unu \citep{PDB4UNU};
    2nrl \citep{PDB2NRL}; 4xe2 \citep{PDB4XE2}; 3chy \citep{PDB3CHY}.
    $^e$Disordering temperature. $^f$Transition temperature.}
\end{table*}
\subsection{Sequence pattern formation is accompanied by a cooperative two-state transition}
We have determined the energy parameters for each of the 15 families of relatively small domains listed in Table \ref{tab:families} by using MC simulations following the procedure described previously \citep{Kinjo2017}. 

\begin{figure}
  \begin{center}
    \includegraphics[width=0.45\textwidth]{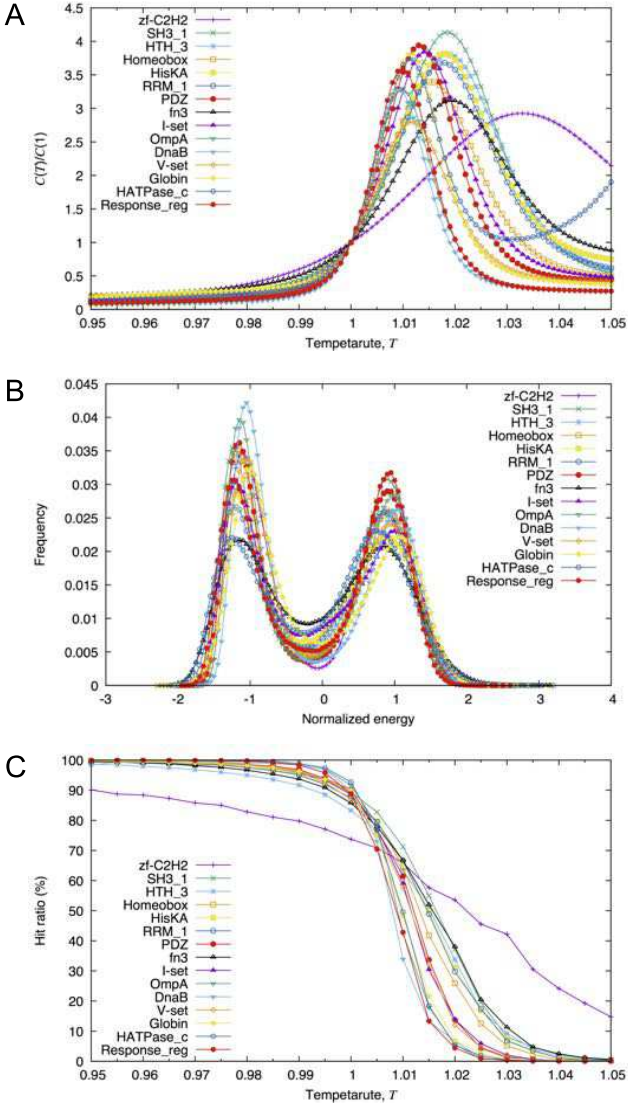}
  \end{center}
  \caption{\label{fig:sim}Results of simulations.
    (A) The specific heat obtained from multicanonical simulations [normalized by $C(T=1)$].
    (B) The energy distributions of grand canonical ensembles at transition temperatures (the energy is normalized by its average and standard deviation at $T=T_m$ for each family).
    (C) The ``hit ratio'', the fraction of the sequences generated by MC simulations at various temperatures that significantly match to the profile HMMs.
  }
\end{figure}

Using the optimized parameters, we run a multicanonical MC simulation \citep{BergANDNeuhaus1992} for each family from which the specific heat,
\begin{equation}
C(T) = \frac{\braket{E^2}_T - \braket{E}^2_T}{T^2},
\end{equation}
was computed as a function of temperature (Fig. \ref{fig:sim}A). For all the families but HATPase\_c, a single peak in the specific heat was observed, indicating the existence of a transition. An increase in the specific heat of HATPase\_c at higher temperatures was due to excessively long insertions. Depending on the family, the peak temperature ($T_m$; see Table \ref{tab:families}) ranged between $T=1.009$ (for DnaB) and $T=1.033$ (for zf-C2H2).  The energy distribution obtained from grand canonical MC simulations at $T_m$ (Fig. \ref{fig:sim}B) showed a clear two-state transition between low energy and high energy states.

In order to confirm if the low-energy sequences are indeed similar to the natural sequences, we compared the profile HMM of each family against the set of sequences generated by the MC simulations. Out of 10,000 sequences generated at each temperature ranging from $T=0.95$ to $T=1.05$, a profile HMM search was performed (using the hmmsearch program \citep{HMMER3}) to identify the sequences that significantly matched the Pfam profile HMM with full sequence E-value $< 0.01$ (Fig. \ref{fig:sim}C). The hit ratio (the fraction of sequences significantly similar to the family's profile HMM) shows the clear two-state transition between natural-like sequences and non-natural-like ($\approx$ random) sequences around the transition temperature $T_{m}$ for each family, and the sequences generated at lower temperatures were indeed more similar to natural sequences. Since these natural-like sequences are expected to fold into the native fold of the domain family but non-natural-like sequences are not, this transition can be regarded as a ``folding transition'' in the sequence subspace.
The hit ratios at $T=1$ (the natural condition) ranged from 73.3\% (for zf-C2H2) and 92.5\% (for HATPase\_c), indicating there were a non-negligible fraction of non-natural-like sequences at the natural condition. In particular, the C2H2-type zinc finger domain (zf-C2H2) consisting of only 23 core sites exhibited a large fraction of non-natural-like sequences at lower temperature.
For all the families examined, the hit ratio was a decreasing function of temperature. For later reference, we defined a ``disordering temperature'', $T_D$, of each family as the lowest temperature at which the hit ratio was less than 1\% (column ``$T_D$'' in Table \ref{tab:families}).

\subsection{Comparison with a model without long-range interactions: An example}
\begin{figure}
  \begin{center}
    \includegraphics[width=0.45\textwidth]{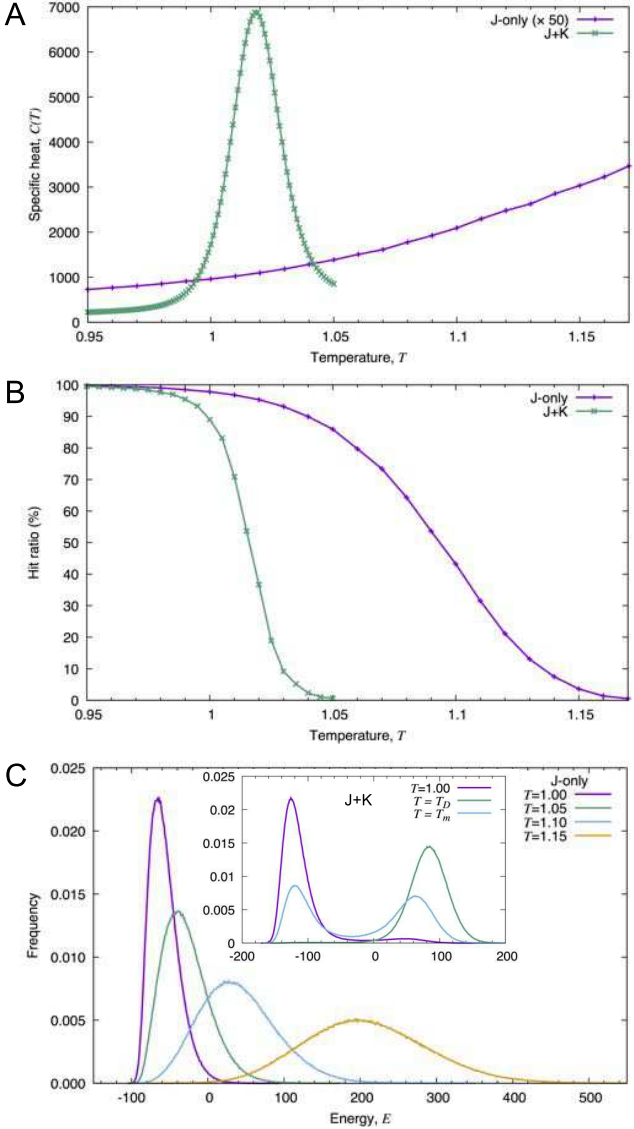}
  \end{center}
  \caption{\label{fig:Jonly} Comparison between the ``J-only'' model and the full ``J+K'' model (c.f., Fig. \ref{fig:sim}) for the SH3 domain (Table \ref{tab:families}).
    (A) Specific heat. The values for the J-only model are multiplied by 50 for clarity.
    (B) Hit ratio. The J-only model had hit ratio less than 1\% at temperatures $T\geq 1.17$ whereas the corresponding temperature (disordering temperature $T_D$) for the J+K model is $T_D = 1.05$ (Table \ref{tab:families}).
    (C) Energy distributions at varying temperatures of the J-only model (the inset shows those of the J+K model). The distributions for the J-only model are always unimodal whereas that for the J+K model at the transition temperature $T_m = 1.019$ is bimodal.}
\end{figure}
The long-range interations incorporated in the LGM are necessary (though not sufficient) for the model to exhibit the cooperative transition.
Conventional sequence models, however, do not include such long-range interactions. In other words, they are essentially one-dimensional models. It is a well-known fact that one-dimensional systems do not exhibit cooperative (phase) transitions at finite temperatures \citep{Landau_Lifshitz_Stat_Phys}. Although theoretically trivial, it may be instructive to demonstrate the absence of the transition in such a system. Within the framework of the LGM, we can simply discard the long-range interations and train a model without the $K$ parameters in the energy function (Eq. \ref{eq:energy0}). In this ``J-only'' model, the $J$ parameters for short-range interactions can be obtained exactly \citep{Kinjo2016}. We trained a J-only model for the SH3 domain family (Table \ref{tab:families}) and compared its specific heat, hit ratio and energy distributions with those of the original ``J+K'' model which includes long-range interactions (Fig. \ref{fig:Jonly}).

The specific heat of the J-only model monotonically increased with temperature (Fig. \ref{fig:Jonly}A) which is in sharp contrast with the J+K model. The upper limit of the temperature range for the J-only model was set to $T=1.17$ because it was only at this temperature the hit ratio became less than 1\% (Fig. \ref{fig:Jonly}B), which is significantly greater than the corresponding disordering temperature for the J+K model ($T_D = 1.05$; Table \ref{tab:families}). Accordingly, the hit ratio decreases rather slowly as the temperature increases. The energy distributions of the J-only model at varying temperatures are unimodal and the mode energy value continuously increases with increasing temperature (Fig. \ref{fig:Jonly}C). This indicates that there is no clear boundary between natural-like sequences and random sequences. 
These results for the J-only model clearly demonstrate the absence of cooperative transition in the model without long-range interactions. It it expected that other conventional sequence models also behave similarly.

In the following, we exclusively use the J+K models.
\subsection{Less conserved sites order at higher temperatures, better conserved sites order at lower temperatures}
\begin{figure}
  \includegraphics[width=0.45\textwidth]{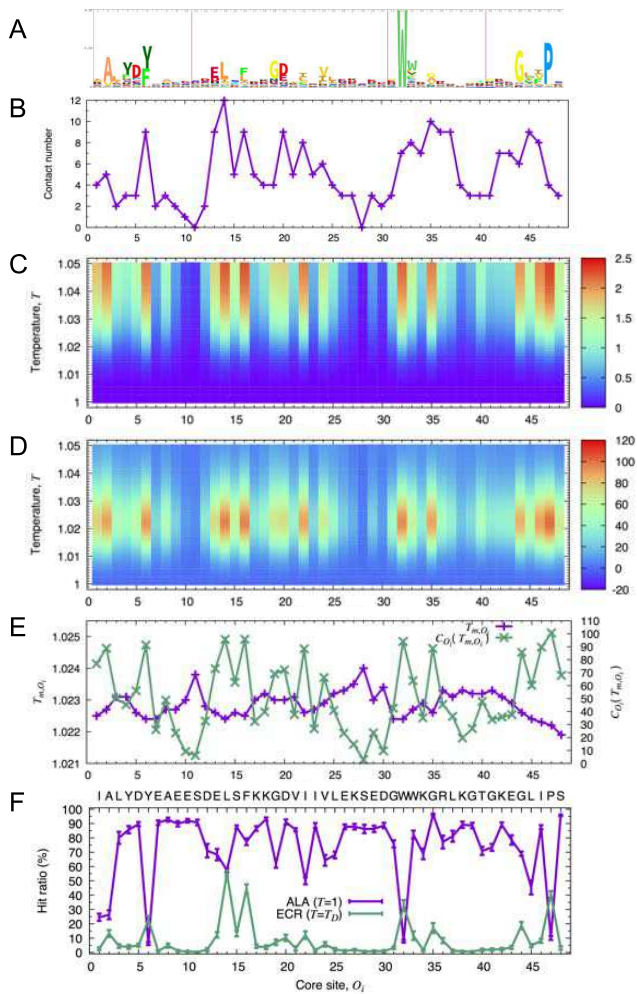}
  \caption{\label{fig:site}
    Site-wise properties for the Pfam SH3 domain family.
    In all the panels, the horizontal axis represents core sites.
    (A) The HMM logo taken from the Pfam site (http://pfam.xfam.org/family/PF00018\#tabview=tab4).
    (B) The contact number of core sites (i.e., the number of non-bonded interactions each site makes with other sites based on the reference structure).
    (C) The divergence map, $D_{O_i}(T)$.
    (D) The differential divergence map $C_{O_i}(T) = \mathrm{d}D_{O_i}(T)/\mathrm{d}T$.
    (E) Site-wise transition temperature $T_{m,O_i}$ (left ordinate) and the differential divergence at the temperature $C_{O_i}(T_{m,O_i})$ (right ordinate).
    (F) Virtual ``mutation'' experiments: ``alanine'' scanning (ALA) and ``enforcedly conserved residue'' scanning (ECR) (see text for the details).
  }
\end{figure}

\begin{figure}
  \includegraphics[width=0.45\textwidth]{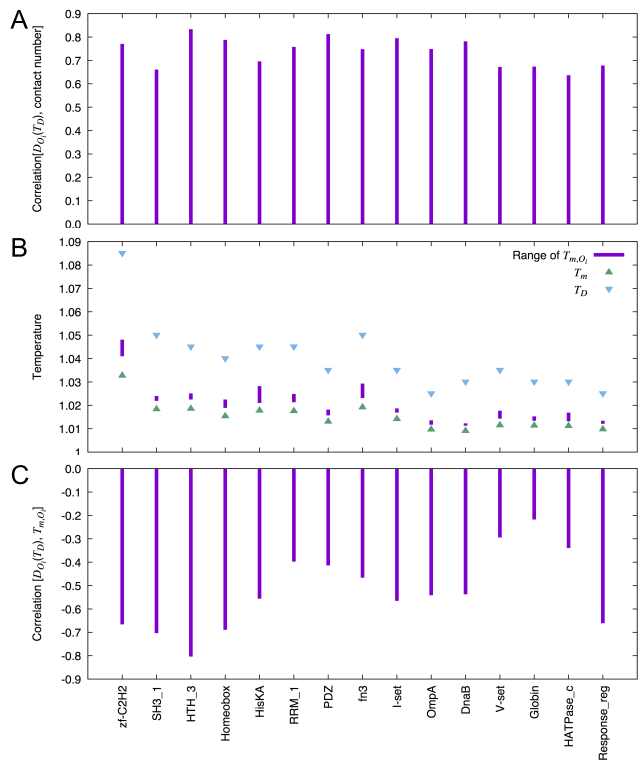}
  \caption{\label{fig:sitecorr}
    Site-wise properties.
    (A) Correlation between site conservation $D_{O_i}(T_D)$ and contact number.
    (B) Range of site-wise transition temperatures $T_{m,O_i}$ together with the global transition temperature $T_m$ and disordering temperature $T_D$.
    (C) Correlation between the site-wise temperatures $T_{m,O_i}$ and site conservation $D_{O_i}(T_D)$.
  }
  
\end{figure}

Next, the transition was characterized in terms of residue distribution at each site over a temperature range.
The ``naturality'' of each core site of simulated sequences at a given temperature was evaluated by the Kullback-Leibler divergence \citep{MacKayInformationTheory} $D_{O_i}(T)$:
\begin{equation}
\label{eq:div}
D_{O_i}(T) = \sum_{a=1}^{21}\braket{n_{(O_i,a)}}_T\log\frac{\braket{n_{(O_i,a)}}_T}{\braket{n_{(O_i,a)}}_{\mathrm{obs}}}
\end{equation}
which measures the difference between the residue distribution $\braket{n_{(O_i,a)}}_T$ of the site $O_i$ at temperature $T$ from the observed residue distribution $\braket{n_{(O_i,a)}}_{\mathrm{obs}}$. An example for the SH3 domain is provided in Fig. \ref{fig:site}C. Lower divergence values imply more natural-likeliness.
By construction, the divergence is nearly 0 for all the sites at $T = 1$, but the values varied largely among different sites for higher temperatures. In general, better conserved sites (c.f., Fig. \ref{fig:site}A) have larger divergence at high temperatures. In the following, we use the $D_{O_i}(T_D)$, i.e., divergence at the disordering temperature $T=T_D$, as a measure of conservation. As expected, sites with a larger contact number \citep{KinjoETAL2005} (i.e., the number of non-bonded contacts a site makes with other sites; see Fig. \ref{fig:site}B for example) tend to be more conserved as can be seen in their correlations (Fig. \ref{fig:sitecorr}A).

Ordering of each site can be quantified via the derivative of the divergence with respect to temperature (c.f., Fig. \ref{fig:site}D):
\begin{eqnarray}
\label{eq:ddiv}
C_{O_i}(T) &=& \frac{\mathrm{d}D_{O_i}(T)}{\mathrm{d}T}\nonumber\\
           &=& \frac{1}{T^2}\sum_{a=1}^{21}\left[\braket{n_{(O_i,a)}E}_{T} - \braket{n_{(O_i,a)}}_{T}\braket{E}_T\right]\nonumber\\
           && \times \log\left[\frac{\braket{n_{(O_i,a)}}_{T}}{\braket{n_{(O_i,a)}}_{\mathrm{obs}}}\right]
\end{eqnarray}
which we call ``differential divergence'' in the following.
Note that this is a weighted sum of covariance between the number density and total energy where the unnormalized and signed weights are $\log[{\braket{n_{(O_i,a)}}_{T}}/{\braket{n_{(O_i,a)}}_{\mathrm{obs}}}]$. Thus, the peaks of this quantity provide the site-specific transition point at which a large change in the residue distribution is accompanied with a large change in total energy.
Precisely, the site-wise transition temperature $T_{m,O_i}$ is defined by
\begin{equation}
\label{eq:tsite}
T_{m,O_i} = \mathrm{arg}\max_{T < T_D} C_{O_i}(T)
\end{equation}
(see Fig. \ref{fig:site}E for an example).
For some sites, this temperature cannot be defined because of the absence of a unique peak for $T < T_D$ (due to lack of non-bonded interactions; this was observed for several sites of the fn3 family).
The range of the site-wise transition temperatures are plotted along with $T_m$ and $T_D$ in Fig. \ref{fig:sitecorr}B. The site-wise temperatures all appeared slightly above the global transition temperature $T_m$, indicating a global (higher order) ordering process between the minimum $T_{m,O_i}$ and $T_m$.
The site-wise transition temperature was found to be negatively correlated with site conservation $D_{O_i}(T_D)$ (Fig. \ref{fig:sitecorr}C) although the correlation was not always significant. Thus, as temperature lowers, local ordering starts from less conserved sites, but these sites have little global effects (smaller values of $C_{O_i}(T_{m,O_i})$; see Fig. \ref{fig:site}E, green line), then it proceeds to increasingly better conserved sites with increasingly greater effects (larger change in total energy). Then, higher-order ordering proceeds down to the global transition temperature, $T_{m}$. 
Interestingly, a nearly perfect correlation ($R \approx 1$) was found between $\{C_{O_i}(T_m)\}$ and $\{D_{O_i}(T_D)\}$. Thus, the transition of each site appears almost completely determined by the conservation patterns.

\subsection{Suppressing key residues inhibits sequence pattern formation under the natural condition: ``Alanine'' scanning}
Although the above analysis of divergence maps hinted some differential roles of individual sites
in determining natural-like sequences, the site-wise transition temperature $T_{m,O_i}$ did not vary
much across the sites. 
In order to explore the role of each site more directly, we performed virtual site-directed mutation experiments by manipulating the chemical potential $\mu$.
First, we performed ``alanine'' scanning experiments. That is, for each core site $O_i$, the chemical potential for alanine (``$\mathtt{A}$'') residue (or glycine ``$\mathtt{G}$'' if the consensus residue was alanine)  was set to a large positive value (namely, $\mu(O_i,\mathtt{A}) = +20$) while other $\mu$'s were set to 0. This effectively fixes alanine residue at the site $O_i$ while allowing other sites to freely adopt residues that are consistent with the perturbation. By performing MC simulations at $T=1$ and applying the hmmsearch program to the sampled sequences, the hit ratio $h_{O_i}^{\text{ALA}}$ (the fraction of natural-like sequences) of the ``alanine mutant'' was determined (c.f., Fig. \ref{fig:site}C, magenta line). This hit ratio was compared to the hit ratio $h_{0}(T=1)$ of the natural condition (c.f., Fig. \ref{fig:sim}C). Namely, we evaluated the ``mutants'' in terms of the following enrichment value:
\begin{equation}
  \text{En}_{O_i}^{\text{ALA}} = h_{O_i}^{\text{ALA}} / h_{0}(T=1).
\end{equation}

\begin{figure}
\includegraphics[width=0.45\textwidth]{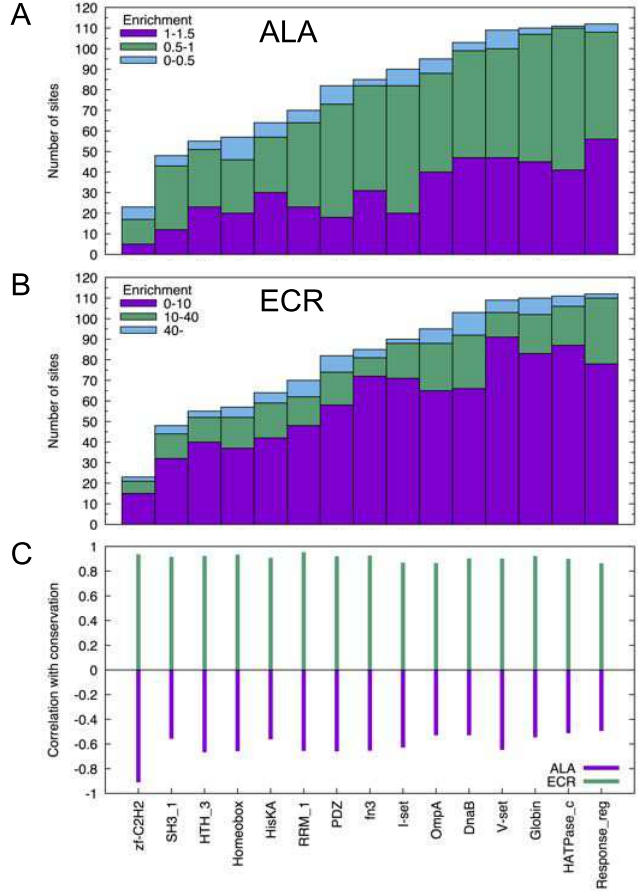}
\caption{\label{fig:mut}
  Virtual mutation experiments. ``ALA'' and ``ECR'' refer to ``alanine mutants'' (at $T = 1$) and ``enforcedly conserved residues'' (at $T = T_D$), respectively. The horizontal axis indicates Pfam families (see the bottom of panel C). A: Histogram of residues with specified enrichment $\text{En}_{O_i}^{\text{ALA}}$ for ALA scanning.
  B: Histogram of residues with specified enrichment $\text{En}_{O_i}^{\text{ECR}}$ for ECR scanning.
  C: Correlation between log hit ratio $\log(h_{O_i}^{\text{ALA/ECR}})$
   and site conservation $D_{O_i}(T_D)$ (see Eq. \ref{eq:div}).
}
\end{figure}

While most of the sites were not largely affected by alanine mutations (the enrichment ranging between 0.5 and 1.5), a few sites exhibited enrichment of less than 0.5 (Fig. \ref{fig:mut}A), that is, the fraction of natural-like sequences for these few mutants was less than a half of that under the natural condition.
The hit ratio of mutants $\{h_{O_i}^{\text{ALA}}\}_{i=1,\cdots,N}$ and a measure of conservation of each site $\{D_{O_i}(T_D)\}_{i=1,\cdots,N}$ showed negative
correlations (Fig. \ref{fig:mut}C; $R < -0.50$ for most families), indicating better conserved sites are affected more severely by the alanine mutation. This result suggests that fixation of some key residues is essential for the global pattern formation of natural-like sequences.

\subsection{Enforcing key residues promotes sequence pattern formation under disordering condition: ``Enforcedly conserved residue'' scanning}
Conversely to alanine scanning, we can enforce the consensus (the most dominant) residue $a_{O_i}^{*}$ of each core site $O_i$ by setting $\mu(O_i,a_{O_i}^{*}) = +20$ (and all other $\mu$'s were set to 0) at the disordering temperature $T = T_D$. With this ``enforcedly conserved residue (ECR)'' mutation at each site, the hit ratio $h_{O_i}^{\text{ECR}}$ was calculated (c.f., Fig. \ref{fig:site}C, green line) and the enrichment with respect to the standard hit ratio at $T_D$
\begin{equation}
  \text{En}_{O_i}^{\text{ECR}} = h_{O_i}^{\text{ECR}} / h_{0}(T=T_D).
\end{equation}
was compared (Fig. \ref{fig:mut}B).
While majority of sites exhibited enrichment of less than 10 fold, a few sites showed more than 40-fold enrichment.
The hit ratio for ECR mutants are even more correlated with site conservation (Fig. \ref{fig:mut}C; $R>0.87$ for all the families).
Thus, enforcing a few well-conserved residues can enhance the sequence pattern formation over the entire sequence under otherwise disordering conditions.

\begin{table*}
  \small
  \caption{\label{tab:mut}Most affected sites by ALA and ECR mutations.}
  \begin{tabular}{lp{6cm}p{6cm}}\hline
    Pfam ID & $En_{O_i}^{\text{ALA}}<0.5$ (ascending order)$^a$ & $En_{O_i}^{\text{ECR}}>40$ (descending order)$^a$ \\\hline
    zf-C2H2&	\ul{C3}, \ul{H19}, \ul{C6}, \ul{H23}, L16, \ul{F10}&	\ul{C3}, \ul{H19} \\
SH3\_1&	\ul{Y6}, \ul{W32}, \ul{P47}, I1, A2&	L14, F16, \ul{P47}, \ul{W32}\\
HTH\_3&	L1, L55, L48, L10&	L15, E30, \ul{R5}  \\
Homeobox&	W47, F48, \ul{R52}, L15, \ul{K56}, A53, \ul{K57}, F19, V44, L33, \ul{N50}&	W47, L15, F48, V44, \ul{R52}  \\
HisKA&	\ul{H11}, R14, E12, L13, P16, \ul{L51}, \ul{L17}&	P16, E12, L13, \ul{H11}, T15\\
RRM\_1&	L1, F46, L15, L6, \ul{F2}, \ul{F43}&	F19, L15, \ul{F43}, V44, L6, A55, V3, A42\\
PDZ&	L79, V30, L44, D48, \ul{L12}, I50, A38, L82, \ul{L16}&	I50, V30, L44, D48, V53, L79, A66, \ul{I70} \\
fn3&	W20, Y33, Y69&	W20, V73, Y33, P3 \\
I-set&	Y71, W34, A35, L58, V90, I60, \ul{P1}, L20&	W34, Y71\\
OmpA&	\ul{R95}, F2, R53, Y1, N94, H36, L49&	R53, V57, A54, F2, \ul{R95}, D38, N46\\
DnaB&	L13, F35, L17, \ul{Y97}&	L13, F35, E9, F44, I43, H40, L17, V12, V62, L84, I47  \\
V-set&	Y88, V108, W33, D84, L16, L75, L109, L106, I77&	Y88, W33, D84, L106, L16, L75  \\
Globin&	\ul{F38}, L24, F28&	F38, F28, \ul{H59}, L24, \ul{F105}, \ul{V63}, W9, G20 \\
HATPase\_c&	\ul{D43}&	\ul{D43}, \ul{N12}, V41, G76, \ul{N16}\\
Response\_reg&	\ul{K100}, \ul{D5}, M53, L48&	\ul{D50}, M53\\
\hline
  \end{tabular}\\
  $^a$Consensus residues with core site position are listed (e.g., ``C3'' indicates the consensus residue C (cystein) at core site 3, $O_3$ for zf-C2H2). Consensus residues involved in intermolecular interactions are marked in \textbf{boldface}.
  PDB ID and references in which intermolecular interactions were confirmed are the following:
  zf-C2H2, 4m9v \citep{PDB4M9V};  SH3\_1, 4hvu \citep{PDB4HVU};
  HTH\_3, 1per \citep{PDB1PER}; Homeobox, 4rdu \citep{PDB4RDU};
  HisKA, 2c2a \citep{PDB2C2A}; RRM\_1, 1po6 \citep{PDB1PO6};
  PDZ, 1b8q \citep{PDB1B8Q}; I-set, 2wp3 \citep{PDB2WP3};
  OmpA, 5eaz \citep{PDB5EAZ}; DnaB, 2vye \citep{PDB4ESV,PDB2VYE};
  Globin, 2nrl \citep{PDB2NRL};
  HATPase\_c, 2wi6 \citep{PDB2WI6}; Response\_reg, 3chy \citep{PDB3CHY}.
\end{table*}

\section{Discussion}
The LGM for studying the ``folding transition'' in the sequence subspace around a given protein family is analogous to G\=o-like models for studying the folding transition in the conformational space for a given native structure \citep{Go1983,ClementiETAL2000}. One major conceptual difference is that while there is a unique native structure as the global energy minimum for a G\=o-like model, there is an \emph{ensemble} of sequences as the global \emph{free energy} minimum for an LGM (Eq. \ref{eq:fe}). This is because in the latter the parameters are optimized using a set of natural sequences at a finite temperature ($T = 1$). This may be considered as a consequence of the asymmetric one-to-many relationship between a protein fold and family sequences.
The design of the LGM assumes a funnel-like free energy landscape of the sequence subspace analogous to that in the conformational space \citep{BryngelsonETAL1995}. Based on simple exact models, it has been suggested that the free energy landscape of the sequence subspace within a family of sequences sharing the same ground-state conformation is indeed funnel-like (``superfunnel'') \citep{Bornberg-BauerANDChan1999,WroeETAL2005}. The present LGM extends the superfunnel concept to include sequences slightly beyond the family boundary.

The parameters of the LGM are determined so that statistics of simulated sequences match that of an observed MSA. This may imply that the sequences sampled by the LGM are biased towards natural sequences. Nevertheless, based on computational protein design experiments, it has been suggested that the volume of sequence space optimal for a protein structure is restricted to a region around the natural sequence\citep{KuhlmanANDBaker2000}. Therefore, the ensemble of sequences sampled by the LGM under the natural condition is expected to be a good approximation for the ensemble of all the sequences, both natural and artificial, compatible with the protein family.

\begin{figure}
  \includegraphics[width=0.45\textwidth]{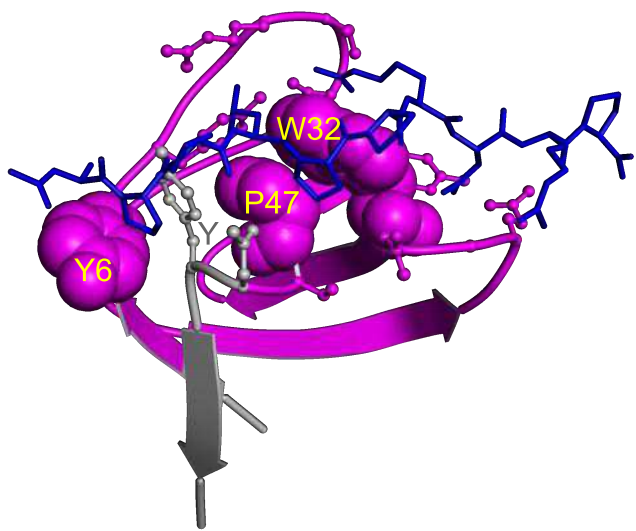}
  \caption{\label{fig:sh3} Chicken src SH3 domain (in magenta and grey ribbon) complexed with a proline-rich peptide (in blue sticks) (PDB: 4hvu \citep{PDB4HVU}).
    The region in magenta indicates residues aligned with the Pfam profile HMM. The residues in the ball-and-stick or spacefill representations are peptide binding residues. The labeled residues in spacefill are among the key residues marked in boldface in Table \ref{tab:mut} and they are part of the ``xP pockets'' \citep{SakselaANDPermi2012}. Note there is a tyrosine residue (Y) in grey which is not aligned with the Pfam profile HMM but forms a xP pocket together with residue Y6. Figure was made with Molmil \citep{Molmil}.
  }
\end{figure}

The existence of a rather sharp two-state transition implies the existence of a clear boundary between family members and non-family members. This transition is made possible by the inclusion of the long-range (non-bonded) interactions in the LGM. Conventional sequence models such as profile HMMs do not (or cannot) incorporate such long-range effects because they are essentially one-dimensional (1D) systems, and therefore, they cannot exhibit such transition \citep{Landau_Lifshitz_Stat_Phys}. Thus, the discrimination between family and non-family members based on the conventional sequence models is necessarily fuzzy (Fig. \ref{fig:Jonly}C), inevitably leading to false positives and false negatives. In fact, this situation may have already biased our understanding of protein families. For example, the C-terminal $3_{10}$ helix and $\beta$ strand ($\beta$5 in \citep{SakselaANDPermi2012}) of the SH3 domain are completely missing in the corresponding Pfam profile HMM (c.f., grey regions in Fig. \ref{fig:sh3}), but these regions not only comprise an integral part of the SH3 fold, but also contain a functionally important tyrosine residue (Fig. \ref{fig:sh3}) involved in peptide binding \citep{SakselaANDPermi2012}. More generally, there are many partial domains found in the MSAs of Pfam and other databases that are shorter than 50\% of the domain model length, and most of them are considered to be alignment and annotation artifacts \citep{TriantANDPearson2015}. The present results suggest that these alignment artifacts are an inherent property of conventional sequence models.
A notable exception to the conventional sequence models is MRFalign \citep{MaETAL2014} based on essentially the same principle as the LGM \citep{MRFbook}. Based on the present discussion, the success of MRFalign in remote homology detection is explained by its ability to set a clear boundary for family members owing to the inclusion of long-range correlations.

In this study, we have used one of the conventional methods, namely the profile HMM, for quantifying natural-likeliness in terms of the ``hit ratio''. However, due to the limitation of profile HMMs just pointed out, such a measure may be inaccurate. In designing artificial WW domain proteins, Socolich et al. \citep{SocolichETAL2005} showed that sequences designed without pairwise correlations were unable to fold whereas all the designed foldable proteins were designed with pairwise correlations (by statistical coupling analysis \citep{LocklessETAL1999}). It has been shown that other statistical models incorporating ``direct'' long-range correlations can discriminate foldable sequences from unfoldable ones to a very good accuracy \citep{BalakrishnanETAL2011,CoccoETAL2017} for the artificially designed WW domains \citep{SocolichETAL2005}. The LGM also incorporates the direct long-range correlations in the form of interacting non-bonded pairs so its energy value is expected to be a good discriminator of foldable proteins.
It is of interest and in principle possible to experimentally measure the foldability of the sequences generated by the LGM.

In the context of molecular evolution, the present results suggest that if some selective pressure is imposed on one or a few key sites, the entire sequence may evolve rather quickly to form a pattern compatible with the family characterized by a unique fold. 
How can such selective pressure act on residues that are to be conserved?
According to recent studies, it appears that protein families have been naturally (and positively) selected so as to stabilize their native folds \citep{MorcosETAL2014,Miyazawa2017}. In fact, most of the key residues listed in Table \ref{fig:mut} comprise the structural core of native folds, suggesting the structural importance of these residues. However, a specific native structure \emph{per se} is unlikely to be subject to natural selection unless it accompanies some advantageous phenotype \citep{Muller2007,NishikawaANDKinjo2014} through molecular functions. Visual inspection of the native structures in Table \ref{tab:families} and their homologs revealed that the key residues often included those involved in intermolecular interactions such as binding of small molecules, polypeptides/proteins or polynucleotides, or phosphorylation (Table \ref{tab:mut}, residues marked in boldface). In the case of the SH3 domain (Fig. \ref{fig:sh3}), for example, the key residues Y6, W32 and P47 (consensus residues according to Fig. \ref{fig:site}A) comprise the ``xP pockets'' essential for the proline-rich peptide binding activity \citep{PDB4HVU,SakselaANDPermi2012}. Thus, although it has been suggested that functionally important residues are not necessarily optimal for structural stability \citep{PandeETAL1994JCP,OtaETAL2003}, some of them play an essential role in cooperatively shaping the overall pattern of family sequences and hence the native fold \citep{YomoETAL1999}. This may also help explain the observation that most of the structural motifs for ligand binding are confined within single protein families \citep{KinjoANDNakamura2009,KinjoANDNakamura2010}. 

\section{Materials and Methods}
\subsection{Data preparation}
All the 15 domain families studied here were selected from the ``top 20'' list of the largest Pfam families (except for the SH3 domain) that had average sequence length of less than 120 residues. Only those of the ``domain'' type were used (i.e., ``family'' and ``repeat'' were discarded). These domain families consist of various folds.
Multiple sequence alignments (MSA) and profile hidden Markov models (HMMs) were downloaded from the Pfam database (version 30.0). The MSAs were based on the ``uniprot'' set when available or on the largest available representative set, otherwise (Table \ref{tab:families}). Aligned sequences that contained more than or equal to 10\% deleted residues on core sites were discarded. A representative structure of each family was assigned based on resolution and alignment quality using the PDBj Mine2 relational database \citep{KinjoETAL2012,KinjoETAL2017,KinjoETAL2018} integrated with the SIFTS \citep{SIFTS2013} resource (Table \ref{tab:families}).

\subsection{Monte Carlo simulations}
In each MC simulation of an LGM model of length $N$, one sweep consists of $N$ and $N-1$ attempted mutations for randomly selected core and insert sites, respectively. For the algorithmic details of the MC simulations of the LGM, refer to \citet{Kinjo2017}.
For the results given in Figs. \ref{fig:sim}B, C, \ref{fig:Jonly} and \ref{fig:mut}, 20 grand canonical MC simulations (under the conditions described in the text) were performed for $10^{6}$ sweeps after $5\times 10^{4}$ sweeps of equilibration, and 10,000 sequences (alignments) were saved every 100 steps for each of the 20 trajectories. For each set of 10,000 sequences resulted from each trajectory, a sequence database was created by removing the ``delete'' symbols from the sequences, against which the hmmsearch \citep{HMMER3} program was applied with the Pfam profile HMM with the E-value cutoff of 0.01 in order to identify significantly similar sequences. From the 20 runs, the average hit ratio and the standard deviation were obtained. 

For the results given in Figs. \ref{fig:sim}A, \ref{fig:site}, and \ref{fig:sitecorr}, first, the density of states was determined by the Wang-Landau method \citep{WangANDLandau2001} by using the energy bins with bin width of 0.2 energy units. The lower bound of the energy range was fixed at a sufficiently low value, and the upper bound was determined by tentative grand canonical MC simulations at a high temperature $T = 1.2$. The Wang-Landau iteration was terminated when the density of states was determined to the precision better than $10^{-7}$.
Next, using thus determined density of states, a multicanonical MC simulation \citep{BergANDNeuhaus1992} was performed until all the energy bins are filled with at least 2,000 counts (sequences were saved at every sweep). From this trajectory, the specific heat and residue distributions at various temperatures were obtained by the reweighting method \citep{NewmanANDBarkema}.

\section{Acknowledgements}
The author thanks Ken Nishikawa, Motonori Ota and Gautam Basu for valuable comments, Eugene I. Shakhnovich for providing some references.




\section*{References}
\bibliographystyle{elsarticle-harv} 
\bibliography{mypaper,refs}

\end{document}